\documentclass[11pt]{article}

\usepackage{amssymb,amsmath,amsfonts}
\usepackage{latexsym}
\usepackage[totalwidth=17cm,totalheight=23.8cm]{geometry}

\usepackage{hyperref}

\newcommand{\be}{\begin{eqnarray}}
\newcommand{\ee}{\end{eqnarray}}

\let\a=\alpha \let\b=\beta  \let\g=\gamma  \let\d=\delta
        \let\l=\lambda
\let\m=\mu    \let\n=\nu          \let\r=\rho
\let\s=\sigma      
    
\let\G=\Gamma \let\D=\Delta   \let\L=\Lambda 
    \let\S=\Sigma     
\let\O=\Omega  \let\eps=\varepsilon

\let\==\equiv

\newcommand{\tr}{{\rm tr}}

\newcommand{\half}{\tfrac{1}{2}}

\newcommand{\cL}{\mathcal{L}}

\newcommand{\cb}{\bar{c}}

\begin{document}

\title{Pure connection gravity at one loop: Instanton background}

\vspace{0.7cm}

\author{Kai Groh, Kirill Krasnov and Christian F.~Steinwachs }

\date{v2: June 2013}

\maketitle

\vspace{-1cm}
\begin{center}
{\it School of Mathematical Sciences, University of Nottingham, Nottingham, NG7 2RD, UK}
\end{center}
\vspace{0.1cm}
\begin{abstract} In the ``pure connection'' formulation General Relativity becomes a particular diffeomorphism invariant ${\rm SL}(2)$ gauge theory. Using this formalism, we compute the divergent contributions to the gravitational one-loop effective action. Calculations of the on-shell effective action simplify greatly if one specialises to an instanton background where only the anti-self-dual part of the Weyl curvature is non-vanishing. One of the most striking features of the connection formulation is that the (linearised) Euclidean action has a definite sign, unlike in the metric case. As in the metric GR, we find the logarithmically divergent contribution to consist of the volume and Euler character terms, but the arising numerical constants are different. However, the difference between the two results turns out to be always {\it an integer}. We explain this by noting that at one loop the connection and metric based quantum theories are closely related, the only difference being in a finite number of scalar modes.
\end{abstract}


\section{Introduction}

General Relativity (GR) admits many reformulations that are equivalent to Einstein's metric formulation at the classical level. The most widely known are the first order formulations: the Palatini formalism where the affine connection becomes an independent field, as well as the tetrad Einstein-Cartan theory. The later is only equivalent to the metric GR in vacuum, while in the presence of matter with spin the connection acquires torsion.  Then there are gauge-theoretic formulations where the tetrad and the spin connection become parts of a larger connection field, see e.g. \cite{Blagojevic:2012bc} for a recent comprehensive review. There is also a geometrically beautiful formulation due to Plebanski \cite{Plebanski:1977zz}, which is based on the notion of self-duality.

All above reformulations work with more independent fields compared to the metric based GR. The latter thus remains the most economical formulation in that it deals with only one dynamical field -- the spacetime metric.  It is one of the main reasons that physicists use the metric GR in practice. For other formulations the equivalence with GR is shown by ``integrating out'' the extra fields present by solving their (algebraic) field equations and substituting the solution back into the action. However, one can also proceed in the opposite direction and, starting from some first order formulation of gravity with its independent metric and connection fields, integrate out the metric. Applying this strategy to e.g. the Palatini formulation leads to a reformulation of GR that is almost as old as GR itself. Nowadays it is known as Eddington's theory, see \cite{Eddington:1987tk}, but it was also known to and studied by Einstein himself, see \cite{Einstein}.

In all known reformulations GR is also perturbatively non-renormalisable, which is manifest in particular in the negative mass dimension of the constant that measures the strength of interactions of gravitons. However, in spite of all reformulations sharing this ``flaw'' of the metric GR, there is no reason to expect that they all lead to exactly the same quantum theory. In this paper we will study precisely such an example of a classically equivalent formulation of GR that behaves rather differently at the quantum level.

The purpose of this paper is to begin to study quantum effects in a reformulation of GR that is similar in spirit to that of Eddington. To obtain this reformulation one starts with the Plebanski formulation \cite{Plebanski:1977zz}, see also \cite{Krasnov:2009pu} for a more recent account. This is a first order formulation, with a two-form field valued in the Lie algebra ${\mathfrak sl}(2)$ carrying the information about the metric, as well as ${\rm SL}(2)$ connection as another independent variable. If this connection is integrated out, one recovers the metric GR Lagrangian (plus an imaginary topological term). If however, as in the passage to the Eddington's formulation, instead the metric-like two-form field is integrated out, one obtains what we refer to as the ``pure connection'' formulation \cite{Krasnov:2011pp}. It is quite similar to the Eddington's formulation because it only makes sense for gravity with a non-zero cosmological constant, with a factor of $1/\Lambda$ in front of the action. Another similarly to Eddington's theory is that the spacetime metric emerges as a derived concept. It is constructed from the curvature of the connection field. However, in contrast to Eddington's formulation, there are much fewer components of the field that the action depends on. Thus, it turns out to provide a very economical description of gravitons, even more economical than the metric GR, see \cite{Krasnov:2011up,Delfino:2012zy} for more details and \cite{Krasnov:2012pd} for a review.

Once GR is reformulated as a diffeomorphism invariant ${\rm SL}(2)$ gauge theory it becomes clear that there is not just one, but instead a large family of theories of interacting gravitons. Then the Lagrangian of General Relativity corresponds to just a special point in the large space of diffeomorphism invariant ${\rm SL}(2)$ gauge theories, all describing just two propagating polarisations of the graviton, see  \cite{Krasnov:2011up}.  It is then tempting to conjecture \cite{Krasnov:2006du} that this family of theories is closed under renormalisation, providing a realization of the asymptotic safety scenario of quantum gravity, see \cite{Weinberg:2009bg} for a recent account. The most direct way to probe this conjecture of closedness under renormalisation is to compute the one-loop counterterms. If it is found that they are of the same form as already present in the tree-level Lagrangian, one could attempt to generalise this to higher loops. If, on the other hand it is found that already at one loop one generates terms not present in the original Lagrangian the conjecture is disproved.

A very efficient technique for performing one-loop computations is the combined use of the background field method and the heat kernel expansion. One expands a given action around an arbitrary background and then uses the heat kernel methods \cite{Vassilevich:2003xt} to compute the regularised determinant of the arising differential operator. It turns out that for a general point in our theory space, and for a general background, the arising differential operator, while second order in derivatives, is too complicated for any simple application of the heat kernel technique (in particular, it is not of a ``minimal'', i.e. Laplace type). So, in this paper, as a first step towards studying the quantum behaviour of diffeomorphism invariant ${\rm SL}(2)$ gauge theories, we perform the one-loop computation for a very special class of backgrounds -- the gravitational instantons. Instantons are defined as the solutions of the Euclidean field equations for which only a chiral half of the Weyl curvature (with our conventions anti-self-dual) is non-zero. It turns out that the action linearised around an instanton background is exactly the same (modulo an overall normalisation factor) for any point in the theory space. Thus, all  diffeomorphism invariant ${\rm SL}(2)$ gauge theories can be treated in one go. In particular, our results are applicable to GR reformulated in this language.

For metric GR, the result of the one-loop computation is known for an arbitrary background.
In this paper, we instead treat an arbitrary member of the above mentioned class of connection theories, but restricted to an instanton background. As discussed in \cite{Krasnov:2009ik}, a general member of this class of theories can be thought of as GR corrected by an infinite set of curvature invariants, when reformulated in the metric language. Thus, such a reformulation when linearized around an arbitrary background would result in a very complicated differential operator.
Therefore a one-loop computation for a general point in this theory space is much more difficult than for GR. On an instanton background, however, most of these difficulties disappear, and the one-loop computation becomes much more feasible. As we shall see, the effort of this computation is even less than the analogous one in the case of metric GR. This simplicity is owed to the fact that in the pure connection formulation a different set of representations of the Lorentz group is propagating as compared to metric GR. In the pure connection formulation, the propagating part of the field transforms in a single irreducible representation. In the metric case, on the other hand, one has to deal both with symmetric tracefree tensors, as well as with scalars.

We find that the final result for the one-loop corrections to the effective action in the connection case is different from the one in the metric theory. In particular, the sign in front of one of the two arising terms in the connection formulation is opposite to that in the metric GR. In general there is no reason to expect to obtain the same quantum theory from two classically equivalent formulations. Although this is confirmed in our calculation, it turns out that the two results are in fact much closer related than expected a priori. Specifically, it turns out that the difference between the one-loop results in the connection and metric formulations is always an integer, determined by the topological properties of the base manifold. This result suggests the existence of a relation between the two quantum theories.

Furthermore, we find that at the one loop level there indeed exists a relation between the metric and connection based quantum theories. This comes from the fact that both formulations can be obtained by starting from one and the same first order action principle, and then integrating out a different set of variables. Starting with the first order Plebanski action linearised around an arbitrary background, one integrates out the connection to arrive at the linearisation of the Einstein-Hilbert action, which is the starting point for the metric one loop computation. If one instead integrates out the linearised metric, one arrives at the linearisation of the pure connection action. Because all the path integrals that arise in this case are Gaussian, it does not matter which set of fields is integrated out first, which establishes a relation between the corresponding quantum theories, at the one loop level. This formal argument misses only a finite number of scalar modes that are present in the metric formalism, and turn out to be absent from the connection description. It is this discrepancy in a finite number of modes that explains the fact that the difference between the two one loop results is always an integer. This argument also suggests that at higher loop order, where non-linearities start to matter, no simple relation between the two quantum theories should be expected to exist.

Another point worth emphasising is the boundedness property of the (linearised) gravity action in the connection formulation. Indeed, it is well-known that the Euclidean Einstein-Hilbert functional suffers from the so-called conformal mode problem. The problem is that this functional is convex (around e.g. the flat space) in the directions of symmetric tracefree variations. It is, however, concave in the direction of the variations of the trace of the metric tensor (i.e. in the directions that correspond to conformal transformations). This makes the Euclidean gravitational action unbounded from below, which in turn makes the Euclidean path integral ill-defined.\footnote{This problem is usually solved, see e.g. \cite{Hawking:1979ig,Gibbons:1978ac} by analytically continuing the conformal factor to pure imaginary values.} In the present pure connection formulation this conformal mode problem is absent. Thus, for the case of positive scalar curvature, one finds that the linearised action is non-negative, and actually becomes a convex functional with no flat directions after an appropriate gauge-fixing. Thus, the Euclidean path integral in the pure connection formulation is a well-defined Gaussian without the need of any analytic continuation in the field space. As we shall see in more details below, these nice properties of the action functional are related to the fact that, in contrast to metric GR, scalar modes are completely absent in the connection formulation of gravity.

The paper is organised as follows.
In Section \ref{sec:lin} we review the class of pure connection theories, as well as their linearisation around an instanton background, and a convenient gauge-fixing term.
In Section \ref{sec:heat} we perform the heat-kernel computation.
Section \ref{sec:sphere} provides an independent check on the results presented.
Here we compute the same regularised determinant by explicitly counting the eigenfunctions of the relevant Laplace operators on the $S^4$. We conclude with a discussion of the results obtained. The Appendix provides an argument relating the two quantum theories at one loop.

\section{The action linearised on an instanton background}
\label{sec:lin}

The content of this section is mostly a review of material presented in \cite{Krasnov:2011up} and \cite{Delfino:2012aj}. The only difference as compared to these references is that we work with a Euclidean version of the theory, as appropriate for a one-loop effective action calculation.

\subsection{Euclidean metric GR}

We start by reviewing the action formulation of Euclidean General Relativity. The action is the familiar Einstein-Hilbert one:
\be\label{action-GR}
S_{\rm EH}[g]=-\frac{1}{16\pi G} \int d^4x \sqrt{g} (R-2\Lambda).
\ee
The reason for the overall minus sign is the following. In the Euclidean path integral one sums over all metrics weighed with a factor $\exp[-S/\hbar]$. This integral should be Gaussian at least for the physical degrees of freedom of the gravitational field -- the modes of the graviton. Gravitons are described as symmetric trace free transverse perturbations of the metric. If we now concentrate on just the first term (i.e. ignore the cosmological constant for the moment), and consider its linearisation around the flat metric $\delta_{\mu\nu}$, we will get for the first variation of the action
\be
\delta S_{\rm EH} = -\frac{1}{16\pi G} \int d^4x \sqrt{g} (R_{\mu\nu}-\frac{1}{2} \delta_{\mu\nu} R)\delta g^{\mu\nu}.
\ee
Now understanding that $\delta g_{\mu\nu}:=h_{\mu\nu}$ we see that the variation of the inverse metric is minus $h_{\mu\nu}$ with indices raised by the flat metric, i.e. $\delta g^{\mu\nu}=-h^{\mu\nu}$.
Assuming the tracefree condition $\delta^{\mu\nu} h_{\mu\nu}=0$ and transverse condition $\partial^\mu h_{\mu\nu}=0$, it is easy to compute the second variation.
The linearised Ricci simplifies to $R_{\mu\nu} = - \partial^\rho \partial_\rho h_{\mu\nu}$, where we have used the following convention for the curvature: $R_{\mu\nu\rho}{}^\sigma = - 2\partial_{[\mu} \Gamma_{\nu]\rho}{}^\sigma + 2 \Gamma_{\rho[\mu}{}^\alpha \Gamma_{\nu]\alpha}{}^\sigma$.
We thus see that for the graviton modes evaluated on the flat background the second variation of the action becomes positive definite
\be
\delta^2 S_{\rm EH} = \frac{1}{16\pi G} \int d^4x \, (\partial_\rho h_{\mu\nu})^2.
\ee
Thus, at least on these graviton modes the Euclidean path integral becomes the usual Gaussian precisely with the minus sign in (\ref{action-GR}). Recalling that there is also a contribution from the cosmological constant term, we see that for $\Lambda>0$ this term is non-negative.

The well-known problem of the Euclidean Einstein-Hilbert action is that with the choice of the sign as in (\ref{action-GR}) there is also the conformal mode, on which the first, Einstein-Hilbert part of the Euclidean action becomes instead {\it negative} definite. Thus, overall, the Euclidean Einstein-Hilbert action is unbounded from below, which makes the Euclidean path integral ill-defined. Several schemes have been proposed for addressing this problem, see e.g. \cite{Hawking:1979ig} and \cite{Mazur:1989by}. Below we will see that this problem does not arise in the pure connection formulation.

It is also worth discussing the case $\Lambda<0$ in more detail. It is clear that in this case the cosmological term in (\ref{action-GR}) is negative definite. This means that even for transverse tracefree perturbations, which are however not sufficiently strongly varying in space, the second negative definite dominates over the first positive term. Thus, in the case $\Lambda<0$ the linearised Einstein-Hilbert actions is not definite even on physical transverse tracefree modes. In contrast, we will see that the pure connection action for $\Lambda<0$, while not positive definite, is at least of a definite sign -- it is negative definite. Again, this is a much more controlled behaviour than that of the Einstein-Hilbert functional.

\subsection{The class of pure connection theories of gravity}

We consider a general diffeomorphism invariant ${\rm SL}(2)$ gauge theory of the form
\be\label{action}
S[A] = - \int d^4x \, f(\tilde{X}^{ij}) \,,
\ee
where
\be\label{X-def}
\tilde{X}^{ij}  = \frac{1}{4} \tilde{\eps}^{\m\n\r\s} F^i_{\m\n} F^j_{\r\s} \,.
\ee
The quantity $\tilde{\eps}^{\m\n\r\s}$ is a completely anti-symmetric tensor density, and $F^i_{\m\n}=2\partial_{[\mu} A_{\nu]}^i+\epsilon^{ijk}A^j_\mu A^k_\nu$ are curvature two-forms for an ${\rm SO}(3)$ connection $A^i_\mu$. The function $f$ is a gauge-invariant real valued function of a matrix argument.
It is also required to be homogeneous of degree one $f(\alpha \tilde{X}^{ij})=\alpha f(\tilde{X}^{ij})$, so that the integrand has the necessary density weight one. We further assume that the connection $A^i_\mu$ is real, implying that the matrix $\tilde{X}^{ij}$ is real as well.
The choice of the minus sign in front of (\ref{action}) is a matter of convention, chosen to have $f$ be a non-negative function later.

The field equations that follow from (\ref{action}) are, in the form notation
\be\label{feqs}
d_A \left( \frac{\partial f}{\partial \tilde{X}^{ij}} \right) \wedge F^j = 0.
\ee
Here $d_A$ is the covariant derivative with respect to the connection $A^i$, defined as $d_A X^i = dX^i + \epsilon^{ijk} A^j X^k$ for an arbitrary section $X^i$.

\subsection{Introducing the metric}

For a generic field configuration $A^i$, at a general point in our 4-dimensional manifold, the curvature 2-forms $F^i_{\mu\nu}$ are linearly independent. They can be declared to be self-dual with respect to a metric, which in turn defines this metric modulo conformal rescalings, see e.g. \cite{Dray} for a discussion most closely related to this context. It can be shown that this metric is Riemannian (i.e.~of signature $++++$) when the real symmetric matrix $\tilde{X}^{ij}$ is definite. Let us assume that $A^i$ is chosen so that this is the case. Then we complete the definition of the metric by defining the volume form via
\be
\frac{\Lambda}{8\pi G} ({\rm vol})=f(\tilde{X}) d^4x.
\ee
Here $\Lambda/G$ has dimensions $1/L^4$, so that both sides of this equality are dimensionless.\footnote{We work in units in which $\hbar=1$, i.e.~coordinates have the dimensions of length, and the connection has dimensions of the inverse length.}
The quantity $\Lambda$ will later become identified with the cosmological constant, and on-shell with the scalar curvature.
The quantity $G$ is the Newton's constant.

Let us now discuss the equivalence with GR.
As is shown in \cite{Krasnov:2011pp}, when choosing the function $f$ in the action (\ref{action}) as
\be\label{f-GR}
f_{\rm GR}(\tilde{X}):=\frac{1}{16\pi G \Lambda} \left( {\rm Tr}\sqrt{\tilde{X}}\right)^2,
\ee
connections $A^i$ satisfying (\ref{feqs}) give rise (by the construction explained in the previous paragraph) to Einstein metrics $R_{\mu\nu}=\L g_{\mu\nu}$ with non-zero scalar curvature.
We note that in the case $\L>0$ of positive scalar curvature the function $f_{\rm GR}$ is non-negative.

The on-shell equivalence to (\ref{action-GR}) can also be seen by evaluating the pure connection formulation action on some simple manifold, e.g.~$S^4$.
In the case of (\ref{action-GR}) this gives
\be\label{EH-on-shell}
S_{\rm EH}[S^4] = -\frac{\Lambda}{8\pi G} \int d^4x \sqrt{g} .
\ee
Bearing in mind the definition of the metric, it is not hard to see that precisely the same value is attained on-shell by (\ref{action}) with the choice $f=f_{\rm GR}$ given in (\ref{f-GR}).

For other choices of $f$ the theory (\ref{action}) continues to describe just two propagating polarisations of the graviton. Perturbatively, when expanded around a constant curvature background, its excitations can be identified as gravitons, see \cite{Krasnov:2011up}. These ``deformations of GR'' can be interpreted in the metric language, see \cite{Krasnov:2009ik}. This is done by introducing a two-form field that encodes the metric directly, and then integrating out the connection. As a result of this procedure one gets a functional of the metric that starts with the usual Einstein-Hilbert term, and continues as a series in curvature invariants. The coefficients in front of different curvature invariants are encoded in the function $f$. One can also deduce the metric interpretation of a general point in our theory space working directly in the connection formulation. This is done by studying the corresponding graviton scattering amplitudes, as in \cite{Delfino:2012aj}.

\subsection{Instanton solutions}

In this paper we specialise to an instanton background. These are connections whose curvature $F^i$ satisfies
\be\label{inst}
F^i\wedge F^j \sim \delta^{ij}.
\ee
Thus, for these connections the quantity $\tilde{X}^{ij}$ as defined by (\ref{X-def}) is a multiple of the identity matrix. Note that such a connection satisfying (\ref{inst}) is automatically a solution of the field equations (\ref{feqs}) for any function $f$.
Indeed, when $\tilde{X}^{ij}\sim \delta^{ij}$ the matrix of first derivatives $\partial f/\partial \tilde{X}^{ij}$ is also a constant multiple of the identity matrix, solving (\ref{feqs}).
Moreover, for any $f$ the metric defined by such $A^i$ is an anti-self-dual Einsteinian metric, as can be seen e.g.~by referring to the Plebanski's formulation of GR \cite{Plebanski:1977zz}. This interpretation of instantons was also noted in \cite{Capovilla:1990qi}, and more recently in \cite{Fine}.
For such a metric only the anti-self-dual part of the Weyl curvature tensor is (possibly) non-vanishing.
The simplest example is the metric on $S^4$, for which all of the Weyl curvature vanishes.

\subsection{Linearised action on an instanton background}

The general action (\ref{action}) can be linearised around an arbitrary background connection $A^i_\mu$. We denote the connection perturbation by $a_\mu^i$. The linearised action (defined as half the second variation) is given by
\be\label{sec-var}
S^{(2)} = -\frac{1}{2} \int \frac{\partial^2 f}{\partial \tilde{X}^{ij}\tilde{X}^{kl}} \Big(\tilde{\eps}^{\m\n\r\s} F^{i}_{\m\n} d_\r a^{j}_\s \Big) \Big( \tilde{\eps}^{\a\b\g\d} F^{k}_{\a\b} d_\g a^{l}_\d \Big) \\ \nonumber
+ \frac{\partial f}{\partial \tilde{X}^{ij}} \Big( 2\tilde{\eps}^{\m\n\r\s} d_\m a^i_\n d_\r a^j_\s + \tilde{\eps}^{\m\n\r\s} F^i_{\m\n} \eps^{jkl} a^k_\r a^l_\s \Big).
\ee
Here, $d_\mu$ is the covariant derivative $d_\mu X^i=\partial_\mu X^i +\epsilon^{ijk} A_\m^j X^k$ with respect to the background connection $A_\mu^i$.

When restricted to an instanton background with $\tilde{X}^{ij}\sim\delta^{ij}$, the linearisation (\ref{sec-var}) simplifies significantly.
As remarked above, in this case the matrix of first derivatives $\partial f/\partial \tilde{X}^{ij}$ is a constant multiple of $\delta^{ij}$.
Integrating by parts in the first term in the second line of (\ref{sec-var}) we find that it cancels the second term precisely.
Thus, only the first line in (\ref{sec-var}) survives in this case.

To rewrite the linearised action in a more manageable form, we explicitly introduce the metric as defined by the background connection. This is most conveniently done by introducing a basis of self-dual two-forms $\Sigma^i_{\mu\nu}$:
\be\label{self-duality}
\frac{1}{2} \epsilon_{\mu\nu}{}^{\rho\sigma} \Sigma^j_{\rho\sigma} = \Sigma^j_{\mu\nu}.
\ee
These two-forms are defined so as to satisfy the following algebra\footnote{This algebra is easily checked for the flat space quantities $\Sigma^i=dt\wedge dx^i + (1/2) \epsilon^{ijk} dx^j \wedge dx^k$.}
\be\label{algebra}
\Sigma^i_\mu{}^\rho \Sigma^j_{\rho\nu} = -\delta^{ij} g_{\mu\nu} - \epsilon^{ijk} \Sigma^k_{\mu\nu},
\ee
as well as the identity
\be\label{ss}
\Sigma^{i\mu\nu} \Sigma^{i\rho\sigma} = g^{\mu\rho} g^{\nu\sigma} - g^{\mu\sigma} g^{\nu\rho} + \epsilon^{\mu\nu\rho\sigma} =: 4P^{+\mu\nu\rho\sigma},
\ee
which follows from the fact that the quantity on the left-hand-side must be a multiple of the self-dual projector.
In view of (\ref{inst}), we can write for the curvature two-forms
\be\label{curv}
F^i_{\mu\nu} = -\frac{\Lambda}{3} \Sigma^i_{\mu\nu}.
\ee

To verify the sign in front of $\Sigma^i_{\mu\nu}$ in (\ref{curv}), we explicitly compute the connection $A^i$ for the 4-sphere. Thus, let us take the metric in the form $ds^2 = d\theta^2 + \sin^2(\theta) d\Omega^2$, where $d\Omega^2=d\eta^2 + \sin^2(\eta) d\xi_1^2+\cos^2(\eta)d\xi_2^2$ is the metric on the 3-sphere in the Hopf fibration coordinates. The self-dual 2-forms for this metric are
\be\nonumber
\Sigma^1&=&\sin(\theta) d\theta \wedge d\eta + \sin^2(\theta)\sin(\eta)\cos(\eta) d\xi_1 \wedge d\xi_2, \\
\Sigma^2&=&\sin(\theta)\sin(\eta)d\theta \wedge d\xi_1+ \sin^2(\theta) \cos(\eta) d\xi_2\wedge d\eta, \\ \nonumber
\Sigma^3&=&\sin(\theta)\cos(\eta)d\theta\wedge d\xi_2+ \sin^2(\theta) \sin(\eta) d\eta \wedge d\xi_1.
\ee
The ${\rm SO}(3)$ connection $A^i$ satisfying $d_A \Sigma^i=0$ is given by
\be
A^1=\cos(\theta) d\eta, \quad A^2=\cos(\theta)\sin(\eta) d\xi_1+\sin(\eta) d\xi_2, \quad
\quad A^3=\cos(\theta)\cos(\eta) d\xi_2+\cos(\eta) d\xi_1.
\ee
Its curvature is given by $F^i=-\Sigma^i$, which justifies the sign in (\ref{curv}).

With this choice of the metric, the matrix $\tilde{X}^{ij}$ is given by
\be\label{X-inst}
\tilde{X}^{ij} = \frac{2\sqrt{g} \,\Lambda^2}{9} \delta^{ij}.
\ee
It is now easy to see that the action (\ref{action}) in the case of $f$ given by its GR form (\ref{f-GR}), when evaluated on an instanton, agrees precisely with (\ref{EH-on-shell}).

It is convenient to redefine the matrix $\tilde{X}^{ij}$ so that the new matrix
\be
X^{ij}:= \frac{9 \tilde{X}^{ij}}{2\sqrt{g} \Lambda^2},
\ee
becomes precisely the identity matrix on the background.
Since the function $f$ is homogeneous in $\tilde{X}^{ij}$, we can freely pass to a function of $X^{ij}$ instead. The linearised action will then contain derivatives of this new function (that we shall still call $f$) with respect to $X^{ij}$. With these choices made, the linearised Lagrangian simplifies to
\be\label{lin-L-1}
{\cal L}^{(2)} = - \frac{\partial^2 f}{\partial X^{ij} \partial X^{kl}} (\Sigma^{i\mu\nu} d_\mu a_\nu^j)(\Sigma^{k\rho\sigma} d_\rho a_\sigma^l).
\ee
Here the matrix of second derivatives is to be evaluated at $X^{ij}=\delta^{ij}$. For the function $f = f_{\rm GR}$ defined in (\ref{f-GR}), the matrix of second derivatives at the identity is given by
\be
\frac{\partial^2 f_{\rm GR}}{\partial X^{ij} \partial X^{kl}} = - \frac{3M_p^2}{2\Lambda} P_{ijkl},
\ee
where $M_p^2=1/16\pi G$ and
\be
P_{ijkl}:= \delta_{i(k}\delta_{l)j} - \frac{1}{3}\delta_{ij}\delta_{kl}
\ee
is the projector on the spin 2 representation of ${\rm SO}(3)$.\footnote{See \cite{Delfino:2012aj} for details of this calculation.}
We see that for the case of GR with positive scalar curvature the matrix of second derivatives is {\it negative definite}. As we shall see shortly, the whole linearised action is then {\it positive definite} (or, more precisely speaking positive semi-definite before the gauge-fixing).

Let us now consider the form of matrix of second derivatives of $f$ for an arbitrary $f$. A simple argument, based on the homogeneity of $f$, then shows that the matrix of second derivatives of $f$ (on the identity matrix) is necessarily proportional to $P_{ijkl}$. Indeed, the homogeneity property can be encoded into
\be
\frac{\partial f}{\partial X^{ij}} X^{ij} = f.
\ee
Differentiating this with respect to $X^{kl}$ implies
\be
\frac{\partial^2 f}{\partial X^{ij} \partial X^{kl}} X^{ij} = 0.
\ee
Evaluated on $X^{ij}=\delta^{ij}$, one immediately concludes that
\be
\frac{\partial^2 f}{\partial X^{ij} \partial X^{kl}} = -\frac{g}{2} P_{ijkl},
\ee
where $g$ is a constant, depending on the function $f$ chosen.
The minus sign in this formula is chosen so that $g$ equals $3M_p^2/\Lambda$ for $f_{\rm GR}$, and is positive for positive $\Lambda$. In what follows we shall assume that we work with functions $f$ such that $g$ is positive.
Thus, we arrive at the linearised Lagrangian
\be\label{lin-L*}
{\cal L}^{(2)} = \frac{g}{2} P_{ijkl} (\Sigma^{i\mu\nu} d_\mu a_\nu^j)(\Sigma^{k\rho\sigma} d_\rho a_\sigma^l).
\ee
It is a significantly simpler linearised Lagrangian than one would obtain in the usual metric formulation, even prior to any gauge-fixing.
When $g>0$ (which is the case for GR with positive $\Lambda$), the Lagrangian is a complete square and is thus non-negative, in contrast to what happens in the metric formulation. For $\Lambda<0$ we get $g<0$ and the linearised functional is non-positive. We will further discuss these properties after we perform the gauge-fixing.

\subsection{Gauge-transformations}

Of course the invariance of (\ref{lin-L*}) under both diffeos and gauge rotations follows from the properties of the action (\ref{action}), but it is instructive to see how they are realised at the linearised level.
The action of the diffeomorphisms on the (infinitesimal) connections $a_\mu^i$ can be conveniently defined as
\be\label{diff-1}
\d_\xi a^i_\m =  \xi^\a F^i_{\a\m}  \,,
\ee
where $\xi^\mu$ is the vector field that generates the transformation.
In this formula the usual Lie derivative of the background connection is corrected by a particular gauge transformation to render an expression that does not contain derivatives of the generator $\xi^\mu$. The transformation rule (\ref{diff-1}) is valid for any background. On our specific background, in view of (\ref{curv}), we can equivalently write:
\be\label{diffeo}
\d_\xi a^i_\m =  \xi^\a \Sigma^i_{\a\m}  \,,
\ee
where we have rescaled the vector field.

The invariance of (\ref{lin-L*}) under diffeomorphisms (\ref{diffeo}) is most conveniently checked by enlarging the derivative operator $d_\mu$ to one compatible with the metric. In (\ref{lin-L*}) the derivative operator $d_\mu$ acting on $a_\nu^i$ only appears with the tensor indices anti-symmetrised $d_{[\mu} a_{\nu]}^i$.
Therefore, it can be enlarged for free to a derivative operator that also acts on the Lorentz indices and is metric compatible (and torsion-free).
This extended operator, which we will continue to call $d_\mu$, satisfies
\be\label{d-Sigma}
d_\mu \Sigma^{i \mu\nu}=0.
\ee
This property can be derived from e.g.~the Bianchi identity $\tilde{\epsilon}^{\mu\nu\rho\sigma} d_\nu F_{\rho\sigma}^i=0$, with the background condition (\ref{curv}).
When $d_\nu$ is extended to a metric-compatible operator, we can take $\epsilon^{\mu\nu\rho\sigma}$  under the derivative operator and then use self-duality (\ref{self-duality}) to obtain (\ref{d-Sigma}).

Once the derivative operator is extended into an operator that acts on both internal and Lorentz indices, we can take $\Sigma^{i\mu\nu}$ under the derivative in (\ref{lin-L*}).
We then see, using (\ref{algebra}), that $\Sigma^{i\mu\nu} a_\nu^j$ transforms under diffeomorphisms as
\be
\delta_\xi (\Sigma^{i\mu\nu} a_\nu^j) = \xi^\mu \delta^{ij} + \epsilon^{ijk} \Sigma^{k\mu}{}_\nu \xi^\nu.
\ee
Both terms here are killed by the projector present in (\ref{lin-L*}), showing diffeomorphism invariance of the linearised action.

The gauge transformations act in the usual way
\be\label{gauge}
\d_\phi a^i_\m = d_\m \phi^i \,.
\ee
Using $2d_{[\mu}d_{\nu]} X^i = \epsilon^{ijk} F^j_{\mu\nu} X^k$, the invariance of the linearized action (\ref{lin-L*}) under the gauge transformation (\ref{gauge}) follows.
Replacing $F^i_{\mu\nu}$ by $\Sigma^i_{\mu\nu}$ and using (\ref{algebra}) one easily finds that $\Sigma^{i\mu\nu} d_\mu a_\nu^j$ changes by a term proportional to $\epsilon^{ijk}\phi^k$, which is again killed by the projector, proving the invariance under the gauge rotations.

\subsection{Gauge fixing}

To gauge fix the diffeomorphism invariance, we observe that the transformation (\ref{diffeo}) has a simple geometrical meaning.
Recall that the Euclidean Lorentz group is just the group of rotations ${\rm SO}(4)={\rm SU}(2)\times {\rm SU}(2)/Z_2$.
Therefore, its representations are characterised by a pair of half-integers (spins) characterising the representations of each ${\rm SU}(2)$.
The field $a_\mu^i$ does not form an irreducible representation of the Lorentz group.
Instead, it decomposes into two components that transform as the irreducible $(3/2,1/2)$ and $(1/2,1/2)$ representations, respectively.

The projector on the representation $(3/2,1/2)$ is explicitly
\be
P^{(3,1)}_{\m i \n j} = \frac{2}{3} \big( \d_{ij}g_{\m\n} - \frac{1}{2}\eps_{ijk}\S^k_{\m\n} \big) \,.
\ee
One can simply verify that the diffeomorphisms (\ref{diffeo}) do not act on the $(3/2,1/2)$ component of the connection at all
\be
P^{(3,1)}{}_\m{}^{i \n j} \xi^\a \Sigma_{\a\n}^j =0 \,.
\ee
At the same time, the other irreducible representation present in $a_\mu^i$, i.e. $(1/2,1/2)$ is changed arbitrarily.
As such, it can be set to zero by an action of a diffeomorphism.
This suggests to fix the gauge for diffeomorphisms sharply and set
\be\label{a-proj}
a_\mu^i = P^{(3,1)}{}_\m{}^{i \n j} a_\nu^j,
\ee
or, equivalently
\be\label{diff-gf}
a_\mu^i = \epsilon^{ijk} \Sigma^j_\mu{}^\nu a_\nu^k.
\ee
With this gauge condition, the corresponding ghost term can be dropped as containing no derivatives of the ghost field for $\xi^\mu$.

The linearised Lagrangian (\ref{lin-L*}) is already a function of only the $(3/2,1/2)$ part of the connection. It remains to gauge-fix the other transformations, i.e. the usual gauge-rotations, in such a way that the complete gauge-fixed Lagrangian continues to depend just on this part of $a_\mu^i$. This is achieved by the gauge-fixing condition of the form
\be
d^\mu (P^{(3,1)}{}_\m{}^{i \n j} a_\nu^j) =0.
\ee
As usual, it is more convenient to add this gauge-fixing condition squared to the Lagrangian, with some appropriately chosen coefficient. A simple calculation of the type given in \cite{Delfino:2012aj} shows that the most convenient gauge-fixing term is
\be\label{Lgf}
{\cal L}_{\rm gf} = \frac{3g}{4} (d^\mu (P^{(3,1)} a)_\mu^i)^2.
\ee
Note that this is of the same sign as the linearised Lagrangian (\ref{lin-L*}), and so non-negative for $\Lambda>0$. To see that this is a useful choice, we rewrite (\ref{lin-L*}), using our gauge-fixing condition (\ref{diff-gf}) as
\be\label{gf-alg-1}
{\cal L}^{(2)} = \frac{g}{2} \delta_{ik} \delta_{jl} (\Sigma^{(i\mu\nu} d_\mu a_\nu^{j)}) (\Sigma^{(k\rho\sigma} d_\rho a_\sigma^{l)}).
\ee
We then write, again using (\ref{diff-gf})
\be
\Sigma^{(i\mu\nu} d_\mu a_\nu^{j)} = \Sigma^{i\mu\nu} d_\mu a_\nu^j - \frac{1}{2} d^\mu a_\mu^k,
\ee
and substitute into (\ref{gf-alg-1}).
Together with the identity (\ref{ss}), we find for the second variation of the action
\be\label{L2selfdual}
{\cal L}^{(2)} = \frac{g}{2} \left( 4P^{+\mu\nu\rho\sigma} d_\mu a_\nu^i d_\rho a_\sigma^i - \frac{1}{2} (d^\mu a_\mu^i)^2\right).
\ee
With the anti-self-dual projector being
\be
P^{-\mu\nu\rho\sigma} := \frac{1}{4} \Big( g^{\mu\rho} g^{\nu\sigma} - g^{\mu\sigma} g^{\nu\rho} - \epsilon^{\mu\nu\rho\sigma} \Big),
\ee
we can insert the relation
\be
4P^{+\mu\nu\rho\sigma} = g^{\mu\rho} g^{\nu\sigma} - g^{\mu\nu} g^{\rho\sigma} + 4P^{-\mu\rho\nu\sigma},
\ee
into (\ref{L2selfdual}), where one should note a different order of indices in the last term.
This yields
\be\label{gf-alg-2}
{\cal L}^{(2)} = \frac{g}{2} \left( (d_\mu a_\nu^i)^2  - \frac{3}{2} (d^\mu a_\mu^i)^2 - 4P^{-\mu\rho\nu\sigma} a_\nu^i d_\mu d_\rho a_\sigma^i\right),
\ee
where we have also integrated by parts in the last term.
Using the anti-symmetry of $P^{-\mu\nu\rho\sigma}$, it can be expressed by the commutator
\be
2 d_{[\m} d_{\r]} a_\s^i = \epsilon^{ijk} F^j_{\m\r} a_\s^k + R_{\m\r\s}{}^\a a_\a^i \,.
\ee
The first term here is purely self-dual, and is killed by the anti-self-dual projector present in (\ref{gf-alg-2}).
In addition we have
\be
4P^{-\mu\rho\nu\sigma} R_{\mu\rho\sigma}{}^\alpha  = -2 R^{\nu\alpha},
\ee
where $R_{\mu\nu}:= R_{\mu\alpha\nu}{}^\alpha$ is the Ricci tensor.
With the on-shell condition $R_{\mu\nu} = \Lambda g_{\mu\nu}$ and the choice for the gauge-fixing term (\ref{Lgf}), we obtain
\be\label{lin-gf}
{\cal L}^{(2)} + {\cal L}_{\rm gf} = \frac{g}{2} \left( (d_\mu a_\nu^i)^2 + \Lambda (a_\mu^i)^2\right),
\ee
where $a_\mu^i$ is understood as the projected quantity (\ref{a-proj}).
Remarkably, for positive cosmological constant $\Lambda>0$ we obtain a {\it positive-definite} linearised functional.

An interesting aspect of our result (\ref{lin-gf}) is that immediately implies that there are no infinitesimal Einstein deformations of instantons with positive scalar curvature.
This directly follows from the fact that the critical points of the functional (\ref{action}) with (\ref{f-GR}) are Einstein.
We have seen that its linearization around an instanton background is positive-definite with no flat directions (after the gauge-fixing), and thus there are no neighbouring critical points. This is a difficult result to prove in the metric formulation, see \cite{LeBrun}.\footnote{Yet a more non-trivial result of Hitchin \cite{Hitchin} states that the only compact Euclidean anti-self-dual Einstein metrics with $\Lambda>0$ are $S^4$ and ${\mathbb CP}^2$. This means that in the case $\L>0$ not just neighbouring critical points are absent, there are no other instanton critical points at all.}

When the cosmological constant is taken to be negative it is not immediately clear if the action functional obtained by integrating (\ref{lin-gf}) over the manifold is of a definite sign.
However, from (\ref{lin-L*}) we see that even prior to the gauge-fixing the Lagrangian is always non-positive. Then we have added a non-positive gauge-fixing term, which keeps (\ref{lin-gf}) non-positive. Since for negative scalar curvature instantons there are no infinitesimal Einstein deformations (see \cite{Rollin}, Proposition 4.5.3 for a simple proof), it follows that there are no zero modes of the operator in (\ref{lin-gf}) even for the case of negative curvature $\Lambda<0$ instantons. Thus, overall, we see that even in the case of the negative curvature our gauge-fixed Lagrangian is definite (more precisely negative definite) with no flat directions.

The fact that the gauge-fixed linearised action is always definite makes the Euclidean path integral much easier to define than in the metric case. Indeed, we recall that in the metric case the linearised action is never definite. For $\Lambda>0$ the metric functional is positive definite in the directions of tracefree transverse tensors, and negative definite in the conformal mode direction. Here the problem can be solved by an analytic continuation of the conformal mode to the pure imaginary values. In our approach this problem is absent altogether: The $\L>0$ functional is positive definite. For $\L<0$ the metric functional exhibits even worse non-definiteness problems. Indeed, as we have already discussed, because of the cosmological constant term working in the opposite direction to the kinetic term, the linearised Einstein-Hilbert functional is indefinite even on the tracefree transverse modes. In our case, for $\L<0$ the connection functional is at least definite. One can then define the path integral by analytically continuing the connection field to the pure imaginary values. This choice of the integration contour will be assumed when we perform the one loop calculation below.

To summarise, the operator in the spin 2 sector is
\be\label{delta-21}
\Delta^{(2,1)} := -d^2 + \Lambda,
\ee
where $-d^2\equiv-d^\mu d_\mu$ is the scalar Laplacian.
Here we have switched to a different labeling of the representations of ${\rm SO}(4)$, more convenient for our purposes.
The representation of highest weight $(2,1)$ corresponds to the representation $(3/2,1/2)$ in the usual labeling, as will be explained in more detail in Section \ref{reptheory}.

\subsection{The ghost sector}

The gauge fixing employed above is accompanied by a Faddeev-Popov ghost term reading
\be
\cL_{\rm gh}= \frac{3}{2} \cb^i \Big( d^\m P^{(3,1)}_{\m i \n j} d^\n \Big) c^j = \cb^i \Big( \d_{ij} d_\n - \half \eps_{ijk}\S^k_{\m\n}d^\m \Big) d^\n c^j.
\ee
The overall coefficient of $3/2$ in the first expression is included so that the operator acquires the standard normalization and does not affect any computation.
Computing the commutator of derivatives in the last term, we get
\be
\cL_{\rm gh}=\cb^i \Big( \d_{ij} d^2 - \tfrac{1}{4} \eps_{ilk}\S^k_{\m\n} \eps_{lmj}F^{m\m\n} \Big) c^j = \cb^i\; \Big( d^2 + \frac{2\Lambda}{3} \Big) c^i \,,
\ee
where we have used (\ref{curv}).
The relevant operator in this case is therefore
\be\label{delta-11}
\Delta^{(1,1)} := -d^2 - \frac{2\Lambda}{3}.
\ee
Again, we have denoted it using the highest weight labelling of the corresponding representation of ${\rm SO}(4)$.
In the more familiar ${\rm SU}(2)\times {\rm SU}(2)$ labelling this is the representation $(1,0)$.

\section{The heat kernel calculation}
\label{sec:heat}

We continue by computing the logarithmically divergent part of the one-loop effective action $\G_{\rm 1-loop}$ based on the heat-kernel expansion for the relevant operators (\ref{delta-21}) and (\ref{delta-11}) derived in the last chapter.

\subsection{Heat kernel technology}

Consider a generalized Laplacian of the form $\Delta = - d^2 - E$ acting on a vector bundle with fiber ${\mathcal R}$; here $E$ is some fiber endomorphism, and $d$ is some covariant derivative operator acting on ${\mathcal R}$-valued functions on the manifold. We are interested in computing the regularised determinant ${\rm det}(\Delta)$. Making use of the identity $\log {\rm det}(\Delta)={\rm Tr}\log(\Delta)$ and replacing the logarithm by its integral representation, one can write
\be\label{log-det}
\log {\rm det}(\Delta)= -\int_0^\infty \frac{dt}{t} {\rm Tr}\left( e^{-t\Delta}\right).
\ee
The trace under the integral is known to admit the heat kernel expansion in powers of $t$
\be\label{expn}
{\rm Tr}\left( e^{-t\Delta}\right) = \int d^4x\, \sqrt{g} \frac{1}{(4\pi t)^2} \sum_{n=0}^\infty t^n a_n^{\mathcal R}(E),
\ee
where the coefficients $a_n^{\mathcal R}(E)$ are local expressions constructed from the curvature of $d$, the endomorphism $E$ and their derivatives \cite{Vassilevich:2003xt}.
In four dimensions, the heat kernel coefficient controlling the logarithmic UV divergence is given by
\be\label{heat}
a_2^{\mathcal R}(E){} &=&{\rm Tr}_{\mathcal R}\Big[
\frac{1}{6}d^2 E + \frac{1}{2} E^2 + \frac{1}{6} R E + \frac{1}{12} \O_{\m\n} \O^{\m\n} \\ \nonumber
&& +\frac{1}{30}d^2 R +\frac{1}{72}R^2 -\frac{1}{180}R_{\m\n}R^{\m\n} +\frac{1}{180}R_{\m\n\r\s}R^{\m\n\r\s} \Big]\,.
\ee
Herein, the curvature $\O_{\m\n}$ is defined as the commutator
\be
\O_{\m\n} = [d_\m,d_\n]
\ee
acting on the space of fields over which the trace is to be taken.

\subsection{The one-loop effective action}

The Euclidean path integral is a sum over all field configurations weighed by $\exp{[-S]}$, with the classical action $S$.
The on-shell one-loop effective action $\G$ can be expressed by expanding $S$ around a stationary point $\phi_{cl}$ via the path integral
\be\label{one-loop}
e^{-\Gamma[\phi_{cl}]} :=
e^{-S[\phi_{cl}]} \int {\cal D}\phi \, e^{-\int \phi\Delta\phi} = \left[ {\rm det}(\Delta)\right]^{-1/2} e^{-S[\phi_{cl}]}.
\ee
Herein $\Delta$ is a general differential operator obtained by linearization of the action $S$, depending on $\phi_{cl}$.
The one-loop contribution to the effective action
\be
\Gamma_{\rm 1-loop}= \frac{1}{2} \log {\rm det}(\Delta),
\ee
can be obtained from (\ref{one-loop}) after splitting $\Gamma = S + \Gamma_{\rm 1-loop}$.
This formula is in turn readily evaluated employing the heat kernel expansion (\ref{expn}) with (\ref{log-det}).

The resulting expression for the one-loop contribution to the effective action
\be
\Gamma_{\rm 1-loop} =
-\frac{1}{2(4\pi)^2} \int_0^\infty dt\;t^{n-3} \sum_{n=0}^\infty \int d^4x\,\sqrt{g}\; a_n,
\ee
requires regularization since the integral over proper time $t$ diverges at the lower bound $t\to 0$ corresponding to the UV limit of the theory.
In a mass scale free renormalization scheme, the running of coupling constants is found from the logarithmically divergent part.
In the above expression, this is given as the term with $n=2$.
Defining
\be\label{gamma-def}
\gamma= \frac{1}{(4\pi)^2} \int d^4x\sqrt{g}\; a_2,
\ee
then allows to write the divergent contribution to $\G_{\rm 1-loop}$ as
\be
\Gamma_{\rm 1-loop}^{\rm log}
= -\frac{1}{2} \g \int_0^\infty \frac{dt}{t}
= \g \log \frac{\m}{\m_0}.
\ee
Here we have regularised the proper time integral by introducing cut-offs $t_{max}\sim 1/\mu^2$ and $t_{min}\sim 1/\mu_0^2$.
It follows that
\be\label{beta}
\frac{\partial \Gamma_{\rm 1-loop}}{\partial \log\mu} = \gamma,
\ee
which establishes the interpretation of (\ref{gamma-def}) as the quantity containing all the $\beta$-functions. We also note that from (\ref{expn}) it can be seen that $\g$ is just the regularized number of eigenvalues of $\Delta$, which can be computed independently.

In the case of the operators in the gravitational (\ref{delta-21}) and the ghost sectors (\ref{delta-11}) respectively, we define
\be\label{gammagrav}
\g_{\rm grav} = \frac{1}{(4\pi)^2} \int d^4x\sqrt{g}\; a_2^{(2,1)}(-\Lambda), \\ \label{gammagh}
\g_{\rm gh} = \frac{1}{(4\pi)^2} \int d^4x\sqrt{g}\; a_2^{(1,1)}(2\Lambda/3).
\ee
The total result for $\gamma$ in the pure connection formulation is then
\be
\gamma_{connection} = \g_{\rm grav} -2 \g_{\rm gh},
\ee
taking into account a factor of $-2$ for the complex valued fermionic ghost fields.
In the following, we compute both contributions separately.

\subsection{The gravitational sector}

For the gravitational contributions we have the operator $\D^{(2,1)}$ (\ref{delta-21}) with the endomorphism part given by $E=-\Lambda$.
The trace in the heat kernel coefficient (\ref{heat}) is here taken over the $(3/2,1/2)$, or equivalently $(2,1)$ in the highest weight labeling, representation of the Lorentz group, which is $8$-dimensional.
The only non-trivial aspect of this calculation is the determination of the matrix $\Omega_{\mu\nu}$.
We have
\be
\O_{\m\n}{}_{\rho i}{}^{\sigma j} a_\s^j &=& [d_\m,d_\n] a_r^i , \\ \nonumber
\O_{\m\n}{}_{\rho i}{}^{\sigma j} &=&-\frac{\Lambda}{3}
\eps_i{}^{kj} \Sigma^k_{\m\n} \d_\rho^\sigma + R_{\m\n\r}{}^\s \d_i^j  \,.
\ee
Its square is given by
\be
\O_{\m\n}{}_{\rho i}{}^{\alpha k} \O^{\m\n}{}_{\alpha k}{}^{\sigma j} =
-\frac{8\Lambda^2}{9} \delta_i{}^j\delta_\rho{}^\sigma - \frac{2\Lambda}{3} \epsilon_i{}^{kj} \Sigma^{k\mu\nu} R_{\mu\nu\rho}{}^\sigma - R^{\mu\nu\alpha}{}_\rho R_{\mu\nu\alpha}{}^\sigma.
\ee
Using the fact that on the instanton background $R=4\Lambda$ the result collapses to just two terms
\be
\tr_{(2,1)}\; \O_{\m\n} \O^{\m\n} =
- \frac{32\Lambda^2}{3} - 2 R_{\m\n\r\s}R^{\m\n\r\s}  \,.
\ee
The subscript $(2,1)$ here signifies the fact that the trace is computed over the $(2,1)$ representation of ${\rm SO}(4)$, by inserting the projector $P^{(3,1)}_{\sigma j}{}^{\rho i}$.
We can now substitute everything into the heat kernel formula (\ref{heat}), drop the surface terms $d^2E$ and $d^2R$, and obtain
\be\label{a2-2}
a_2^{(2,1)}(-\Lambda) = -\frac{28}{45} \Lambda^2 -\frac{11}{90} R_{\m\n\r\s}R^{\m\n\r\s}\,.
\ee

\subsection{The ghost sector}

On the space of ghost fields we have the operator $\D^{(1,1)}$ (\ref{delta-11}), with the endomorphism $E=2\Lambda/3$.
The curvature $\O_{\m\n}$ becomes on this field space
\be
\O_{\m\n}{}^{i}{}_{j} c^j =
[D_\m,D_\n] c^{i} =
\eps^{ikj} F^k_{\m\n} c^j \,.
\ee
Here, one evaluates the traced curvature square on the instanton background, yielding
\be
\tr_{(1,1)}\; \O_{\m\n} \O^{\m\n} =
-\frac{8}{3} \Lambda^2 \,.
\ee
The coefficient is then found to be
\be\label{a2-1}
a_2^{(1,1)}(2\Lambda/3) = \frac{107}{45} \Lambda^2 +\frac{1}{60} R_{\m\n\r\s}R^{\m\n\r\s} \,.
\ee

\subsection{Final result and comparison with the metric case}

Combining the results from the gravitational and ghost sectors (\ref{a2-2}) and (\ref{a2-1}), we get the following final result for the logarithmically divergent part of the one-loop counterterm
\be\label{result}
\gamma_{connection}=\frac{1}{(4\pi)^2}\int d^4x \sqrt{g}\left( -\frac{242}{45} \Lambda^2 -\frac{7}{45} R_{\m\n\r\s}R^{\m\n\r\s}\right) \,.
\ee
For comparison, the analogous result in metric gravity is \cite{Christensen:1979iy}
\be\label{result-GR}
\gamma_{metric}=\frac{1}{(4\pi)^2}\int d^4x \sqrt{g} \left(  -\frac{522}{45} \Lambda^2 +\frac{53}{45} R_{\m\n\r\s}R^{\m\n\r\s}  \right) \,.
\ee
We see that the sign in front of the topological Riemann-squared term has changed in the pure connection gravity as compared to the metric gravity.

It is useful to rewrite the above expressions for $\gamma$ in terms of dimensionless quantities.
We recall the topological Euler character, which on-shell is given by
\be\label{chi}
\chi:=\frac{1}{32\pi^2} \int d^4x \sqrt{g} \, R_{\m\n\r\s}R^{\m\n\r\s}.
\ee
On an instanton metric, the volume is related to another topological invariant, the signature $\tau$.
We have the linear combination of $\chi$ and $\tau$
\be\label{tau}
2\chi+3\tau = \frac{\Lambda^2}{6\pi^2} \int d^4x\sqrt{g}.
\ee
This relation can be obtained from the expressions for $\chi$ and $\tau$ in terms of the components of the curvature tensor, see e.g. \cite{Besse}, page 161.

Therefore, we can express the result (\ref{result}) in terms of the invariants $\chi$ and $\tau$ as
\be\label{result-chi-tau}
\gamma_{connection} = -\frac{391}{90}\chi - \frac{121}{20}\tau.
\ee
As it follows from (\ref{tau}), for an instanton of non-zero scalar curvature $\chi>(3/2)|\tau|$.\footnote{This inequality is in fact more general and holds for an arbitrary Einstein manifold of non-zero scalar curvature.}
This implies that
\be\label{gamma-ineq}
\gamma_{connection}<-\frac{14}{45}\chi.
\ee
Given that on any Einstein manifold of non-zero scalar curvature $\chi$ is positive, this further implies that $\gamma_{connection}<0$. Note that the formula (\ref{result}) is only valid on the instanton background with $F^i\wedge F^j\sim\delta^{ij}$, so that it becomes independent of the details of the function $f$, whereas (\ref{result-GR}) is valid for a general Einstein space. So, the expression (\ref{result-chi-tau}) and the result (\ref{gamma-ineq}) also hold only for instanton solutions.

Finally, let us give an expression for the difference between the two results (\ref{result}) and (\ref{result-GR}) for an instanton metric.
In this case we can express the volume via (\ref{tau}) to obtain
\be\label{diff}
\gamma_{connection} - \gamma_{metric} = 2\chi+7\tau.
\ee
We stress that this difference is always an integer, depending only on topological constants.
For example, on the 4-sphere $\chi=2,\tau=0$ and $\gamma_{connection} - \gamma_{metric} = 4$.
On ${\mathbb CP}^2$ we have $\chi=3,\tau=-1$ and the difference takes value $\gamma_{connection} - \gamma_{metric} = -1$.
We will provide an explanation for this result for the case of $S^4,{\mathbb CP}^2$ in the next Section.

\section{Consistency check on $S^4$}
\label{sec:sphere}

We now provide an independent group theoretic verification of the result (\ref{result}) by explicitly counting eigenvalues of the operators (\ref{delta-21}) and (\ref{delta-11}) in the case of the 4-sphere.
We start by reviewing the relevant facts of representation theory of ${\rm SO}(4)$ and ${\rm SO}(5)$, most of which are presented in \cite{Camporesi:1995fb}.

\subsection{Some representation theory}
\label{reptheory}

To count eigenvalues, we need to understand the spectrum of the scalar Laplacian $\Delta=-d^2$ on functions on $S^4$ taking values in an appropriate representation of the Lorentz group ${\rm SO}(4)$. We first recall that since ${\rm SO}(4)= {\rm SU}(2)\times {\rm SU}(2)/Z_2$, its irreducible representations are characterised by a pair of spins (with both spins being either integers of half-integers).
In this parameterisation the fundamental spinor representations are $(1/2,0)$ and $(0,1/2)$.
However, often the so-called highest weight labelling of representations is more convenient as it generalises to an arbitrary ${\rm SO}(n)$.
In this labelling the fundamental representations are $(1/2,1/2)$ and $(1/2,-1/2)$ instead.

The irreducible representations of interest to us here are the second symmetric power of the representation $(1/2,1/2)$, which is the representation $(1,1)$, as well as the third symmetric power of $(1/2,1/2)$ times $(1/2,-1/2)$, which is $(2,1)$. So, we see that the first of these numbers is the total spin, while the second characterises the ``chirality'' of the representation. For instance the usual symmetric tracefree tensors form the representation $(2,0)$, which is not chiral. In contrast, in our description of gravitons a chiral representation $(2,1)$ is used.

The dimension of each representation $\lambda=(s,\kappa)$ is given by the Weyl formula
\be\label{Weyl}
{\rm dim}_\lambda = \prod_\alpha \frac{\alpha\cdot (\lambda+\rho)}{\alpha\cdot \rho},
\ee
and the quadratic Casimir is
\be\label{Casimir}
C_\lambda = (\lambda+\rho)^2-\rho^2.
\ee
Here, $\a$ denotes the positive roots of the corresponding ${\rm SO}(n)$, whereas $\rho$ is half the sum of the positive roots.
In the case of ${\rm SO}(4)$, these are given by
\be
\alpha_{\rm SO(4)} = \{(1,1), (1,-1)\},
\ee
and
\be
\rho_{\rm SO(4)} = (1,0),
\ee
respectively.
Thus, we see that the representation $(2,1)$ is 8-dimensional, while $(1,1)$ is 3-dimensional.
Furthermore, for the two representations of interest we have
\be\label{c4}
C_{(2,1)}=9, \qquad C_{(1,1)}=4.
\ee

We can now study the spectrum of the scalar Laplacian on functions with values in $(2,1)$ and $(1,1)$.
This problem is readily solved by the representation theory of ${\rm SO}(5)$, which is relevant since $S^4={\rm SO}(5)/{\rm SO}(4)$.
Irreducible representations of ${\rm SO}(5)$ can also be described by their highest weight. As for ${\rm SO}(4)$, one needs to specify just two numbers.
The formulas (\ref{Weyl}) and (\ref{Casimir}) are then still valid, and we just need to know the set of positive roots
\be
\alpha _{\rm SO(5)}=\{ (1,0), (0,1), (1,1), (1,-1)\},
\ee
and their half-sum given by
\be
\rho_{\rm SO(5)}=(3/2,1/2).
\ee

Representations of ${\rm SO}(5)$ that are relevant for our problem can be characterised as follows.
Consider a representation $\tau$ of ${\rm SO}(5)$, and its restriction to ${\rm SO}(4)$.
This restriction is typically not an irreducible representation, but can be decomposed into a direct sum of irreducibles of ${\rm SO}(4)$.
The relevant representations are those for which this decomposition contains the given representations $(2,1)$ and $(1,1)$.
To understand which series of representations of ${\rm SO}(5)$ satisfy this criterion we use Theorem 2, page 228 of \cite{Barut:1986dd}.
It is then easy to deduce that the representations $(n+2,2)$ and $(n+2,1)$ when restricted to ${\rm SO}(4)$ both contain $(2,1)$, and that the representation $(n+1,1)$ contains $(1,1)$.
For each series $n=0,1,\ldots$.
Note that the series of representations $(n+1,1)$ is the same as $(n+2,1)$, just starting with different $n$.
Thus, we have to consider only two different series of representations of ${\rm SO}(5)$.
Their dimensions are found via (\ref{Weyl}) to be
\be\label{dims}
{\rm dim}_{(n+2,2)} = \frac{5}{6}(2n+7)(n+6)(n+1), \\ \nonumber
{\rm dim}_{(n+2,1)}= \frac{1}{2}(2n+7)(n+5)(n+2),
\ee
and the values of the corresponding quadratic Casimirs from (\ref{Casimir}) are given by
\be
C_{(n+2,2)}= n^2+7n+16, \\ \nonumber
C_{(n+2,1)}= n^2+7n+12.
\ee

With this information at hand it is easy to deduce the spectra of the Laplacian on the corresponding spaces.
For each series, the spectrum of the scalar Laplacian is given by the quadratic Casimir of the corresponding representation of ${\rm SO}(5)$ minus the quadratic Casimir of the ${\rm SO}(4)$ representation on which the Laplacian acts
\be
w_{\l_{\rm SO(5)}} = C_{\l_{\rm SO(5)}} - C_{\l_{\rm SO(4)}}.
\ee
Recalling (\ref{c4}) we immediately see that the spectrum of the scalar Laplacian on $(2,1)$ consists of two series with
\be\label{w-21}
w_{(n+2,2)} = n^2+7n+7, \qquad
w_{(n+2,1)} = n^2+7n+3,
\ee
while the spectrum on $(1,1)$ consists of a single serie with
\be\label{w-11}
w_{(n+1,1)} = n^2+5n+2.
\ee
In each case, the multiplicity of the eigenvalue is just the dimension (\ref{dims}) of the corresponding ${\rm SO}(5)$ representation.

\subsection{$\zeta$-function regularization}

The number of eigenvalues of the Laplacian on the sphere $N$ can be decomposed into contributions from negative, zero and positive modes
\be
N:=N_{-}+N_{0}+N_{+}.
\ee
In general, $N_{-}$ and $N_{0}$ are finite. However there are infinitely many positive modes, so that $N_{+}$ has to be regularized by
\be
N_{+}:=\underset{s\to 0}{\lim}\, \zeta(s)\,.
\ee
Here the generalized $\zeta$-functions are defined as
\be
\zeta_\l(s) := \sum_{n\in N^{+}}\,{\rm dim}_{\lambda}(n)\cdot \big( w_{\lambda}(n)\big)^{-s}.
\ee
All the $\zeta$-functions appearing for the spectra and their respective multiplicities can be written in the form
\be\label{zeta}
\zeta(s)=\left( \frac{12}{\Lambda}\right)^s \sum_{n=m}^\infty \frac{(2n+1)[ (2n+1)^2-a^2]}{[(2n+1)^2-b^2]^s}.
\ee
As is shown in the Appendix of \cite{Christensen:1979iy} this sum can be regularized for $s\to 0$ to yield
\be\label{zetareg}
\underset{s\to 0}{\lim}\, \zeta(s) = -\frac{7}{120}+\frac{1}{8}b^4-\sum_{n=0}^{m-1}(2n+1)^3- \frac{1}{12}a^2-\frac{1}{4}a^2 b^2+a^2\sum_{n=0}^{m-1}(2n+1).
\ee

\subsection{Counting eigenvalues}

In our case, there are the three spectra (\ref{w-21}), (\ref{w-11}) to consider.
Let us first consider the spectrum of the operator (\ref{delta-21}) acting in the space $(2,1)$.
There are no zero or negative modes in this case.
The spectrum of (\ref{delta-21}) is obtained by taking (\ref{w-21}), multiplied by the squared inverse radius of the sphere (equal to $\Lambda/3$), and adding the shift due to the mass term in (\ref{delta-21}).
We denote the spectra shifted in this way by $w^\L$.
Overall, we get the following two series
\be
w_{(n+2,2)}^{\L} = \frac{\Lambda}{3}(n^2+7n+10), \qquad
w_{(n+2,1)}^{\L} = \frac{\Lambda}{3}(n^2+7n+6).
\ee
The corresponding $\zeta$-functions can be written in the form (\ref{zeta}) and read
\be\label{zeta-12}
\zeta_2(s) =\sum_{n=0}^\infty \frac{\frac{5}{6}(2n+7)(n+6)(n+1)}{[M^2(n^2+7n+10)]^s} = \frac{5}{24} \left( \frac{12}{\Lambda}\right)^s \sum_{n=3}^\infty \frac{(2n+1)[ (2n+1)^2-25]}{[(2n+1)^2-9]^s}, \\ \nonumber
\zeta_1(s)=  \sum_{n=0}^\infty \frac{\frac{1}{2}(2n+7)(n+5)(n+2)}{[M^2(n^2+7n+6)]^s} = \frac{1}{8} \left( \frac{12}{\Lambda}\right)^s \sum_{n=3}^\infty \frac{(2n+1)[ (2n+1)^2-9]}{[(2n+1)^2-25]^s}.
\ee
We note that these are precisely the $\zeta$-functions encountered in the metric GR computation in \cite{Christensen:1979iy}.

For the operator (\ref{delta-11}) acting on the $(1,1)$ representation, after the shift by the mass term, we find the spectrum
\be
w_{(n+1,1)}^{\L} = \frac{\Lambda}{3}(n^2+5n).
\ee
Thus, there is a zero mode with multiplicity $10$ in the spectrum.
For the positive eigenvalues (i.e. modes with $n=1,2,\ldots$) one obtains precisely the same $\zeta_1(s)$ as given in (\ref{zeta-12}).

Using (\ref{zetareg}), the two $\zeta$-functions in (\ref{zeta-12}) evaluate to
\be
\zeta_2 := \lim_{s\to0} \zeta_2(s)= \frac{89}{18}, \qquad \zeta_1 := \lim_{s\to0} \zeta_1(s)=-\frac{191}{30}.
\ee
Thus, the total (regularized) number of eigenvalues of the operator (\ref{delta-21}) is given by
\be\label{zeta-sum-2}
N = N_{+} = \zeta_2+\zeta_1= -\frac{64}{45}.
\ee
This should be compared to the quantity $\g_{\rm grav}$ defined in (\ref{gammagrav}), given by the heat kernel coefficient (\ref{a2-2}).
The volume of the 4-sphere of radius $\sqrt{3/\Lambda}$ is $24\pi^2/\Lambda^2$, and $R_{\mu\nu\rho\sigma}R^{\mu\nu\rho\sigma}= 8\Lambda^2/3$ so that the Euler character given by (\ref{chi}) is $\chi=2$.
Inserting these quantities into (\ref{a2-2}) we reproduce precisely the number in (\ref{zeta-sum-2}).

Similarly, we check the contribution from the ghost sector.
This is given by the sum of the number of zero modes, which is 10, plus the corresponding $\zeta$-function.
This gives
\be
N = N_0 + N_{+} = 10+\zeta_1=\frac{109}{30}.
\ee
Again, comparing this with $\g_{\rm gh}$ defined in (\ref{gammagh}) given by the heat kernel coefficient (\ref{a2-1}), evaluated on the sphere, we find perfect agreement.

\subsection{Interpretation of the difference}

Overall, the $S^4$ computation given in this Section is significantly simpler than the analogous computation for metric gravity in \cite{Christensen:1979iy} in several respects.
First, there are no scalars present, which means that only two instead of three $\zeta$-functions need to be computed.
Moreover, only the operator arising in the ghost sector has zero modes. There are no negative eigenvalue modes in our case. In contrast, both the vector and the scalar parts in \cite{Christensen:1979iy} had negative eigenvalue modes.

The computation presented also helps to understand the difference (\ref{diff}) between the metric GR result and our case. As is explained in \cite{Christensen:1979iy}, on the 4-sphere the metric result (\ref{result-GR}) has the following composition
\be\label{gamma-gr}
\gamma_{metric} &=& (\zeta_2+\zeta_1+\zeta_0)+(6+\zeta_0)-2(5+10+\zeta_1+\zeta_0) \\ \nonumber
&=& \zeta_2 - \zeta_1 -24 = -\frac{571}{45}.
\ee
Our case is very similar and we have the following structure of the final result
\be\label{gamma}
\gamma_{connection} = (\zeta_2+\zeta_1)-2(10+\zeta_1) \\ \nonumber
= \zeta_2 - \zeta_1 -20 = -\frac{391}{45}.
\ee
We note that in (\ref{gamma-gr}) the scalar $\zeta$-functions cancel, and so the only difference between (\ref{gamma-gr}) and (\ref{gamma}) is that in the latter there are no negative eigenvalue modes for the scalars and ghosts
\be\label{conn-metr}
\gamma_{connection}-\gamma_{metric}=2\times 5-6=4.
\ee
Taking into account that the 6 here is composed of 1 scalar Laplacian eigenfunction with eigenvalue $-2\Lambda$, and 5 modes with eigenvalue $-2\Lambda/3$, see \cite{Christensen:1979iy}, and that 5 negative eigenvalue modes in the vector sector also have their origin in the Laplacian on scalars, we see that the difference between the metric and the connection calculations is that the latter is completely free of any scalar Laplacian contributions.
For non-negative eigenvalues this happens automatically already in the metric computation, but the connection calculation in addition disregards the negative eigenvalue modes as well.
This provides a clear way to understand why the difference between the two results is an integer.

So far, for the case of $S^4$, we have understood the integer arising in the difference (\ref{diff}) as being due to negative eigenvalue modes of the scalar Laplacian, which appear only in the metric calculation. The other example that can be treated completely explicitly, and where a similar interpretation of the difference is possible is $CP^2$. Relevant facts about the scalar Laplacian in this case can be found in \cite{Vassilevich:1993yt}. What is relevant for us here is that in the case of $CP^2$ there are no conformal Killing vector fields. So, in the metric calculation, there are no negative eigenvalue modes in the vector sector, and there is just a single negative eigenvalue mode in the scalar sector, the one that corresponds to constant rescalings of the metric. In the case of $CP^2$, it is just this single rescaling mode that is counted by the metric, but is not counted by the gauge-theoretic approach. This explains the fact that for $CP^2$ the difference $\gamma_{connection}-\gamma_{metric}=-1$.

It is tempting to speculate that for an arbitrary instanton the difference (\ref{diff}) counts just the same: it is twice the number of negative eigenvalue modes of the vector Laplacian minus the number of negative eigenvalue modes of the scalar Laplacian. We leave attempts to prove this to further research.

\subsection{Subtleties: zero modes}

Both in the metric and in our gauge-theoretic calculations no special treatment was given to the zero modes. Recall that in both cases there are 10 zero modes, which in the metric case have the interpretation of the 10 Killing vector fields generating the isometries of $S^4$. However, it has been emphasised in the literature, see in particular \cite{Vassilevich:1992rk}, that the zero modes should be given a separate careful treatment. For example, the authors of \cite{Mazur:1989by} argued that the zero modes should be omitted from the path integral completely. The current consensus is that the zero modes do have to be included, but have to be treated non-perturbatively. This can be done in a variety of schemes, see \cite{Vassilevich:1992rk} for a discussion. In the case of $S^4$ the complete one loop analysis that computes the full $\Gamma_{\rm 1-loop}$ and not just the scaling $\partial \Gamma_{\rm 1-loop}/\partial \log\mu$ appeared only in \cite{Volkov:2000ih}. This reference treats the zero modes non-perturbatively, and does find that they contribute to the scaling in precisely the way as the naive analysis based on the heat kernel expansion predicts. Thus, the formula for $\gamma_{metric}$ obtained in \cite{Christensen:1979iy} is confirmed by this more careful analysis.

What this discussion means for our problem is that, in principle, the zero modes that arise in the ghost sector in our case should be given a careful non-perturbative treatment. However, since we are only interested in the scaling $\gamma_{connection}$, it is possible to count these zero modes on exactly the same footing as all other. This is what was done by our calculations above. A non-perturbative treatment of the zero modes then only becomes relevant in case the full one loop partition function $\Gamma_{\rm 1-loop}$ is needed. We leave this to future research.

\section{Discussion}

Firstly, we comment on the close relationship between the results for pure connection (\ref{result}) and metric gravity (\ref{result-GR}).
Given that the connection formulation is only equivalent to the metric GR on-shell, in principle, there was no reason to expect any simple relation between the quantum theories. However, as we have already explained in the Introduction, the fact that both formulations can be obtained from the same first order Plebanski formulation explains why such a relation exists at one loop. We give a detailed explanation of this in the Appendix.

Let us now discuss the difference between the metric and connection calculation results. According to (\ref{beta}), both coefficients in the obtained result for the one-loop effective action in the pure connection formulation (\ref{result}) correspond to the $\beta$-functions for the associated couplings in the original Lagrangian. The $\beta$-function in front of the Euler character term $\chi$ has a different sign as compared to the one in metric GR.
The other $\beta$-function has the same sign, although its interpretation is different.
Let us first discuss the case of the metric GR.

Given the Euclidean on-shell action (\ref{EH-on-shell}), we can write
\be\label{LG1}
\frac{\partial}{\partial \log{\mu}} \left( \frac{\Lambda}{8\pi G}\right) = \frac{58\Lambda^2}{5(4\pi)^2}.
\ee
Here we have cancelled the volume factors and the minus signs on both sides of the equation.
This formula describes the logarithmic running of the dimensionful quantity $\Lambda/G$ with energy, proportional to $\Lambda^2$.

Then, in the absence of matter, one can define the dimensionful metric $\bar{g}=\Lambda g$. When the Einstein-Hilbert Lagrangian is rewritten in terms of $\bar{g}$, it only contains the dimensionless combination $\Lambda G$, appearing as the prefactor $1/\Lambda G$ in front of the action. This makes it obvious that only the dimensionless combination $\Lambda G$ has a meaningful running in a pure gravity theory.
Since $\Lambda$ appears as a length scale of the background, we can choose it as a constant and rewrite (\ref{LG1}) as
\be\label{RG-gr}
\frac{\partial\, (\Lambda G)}{\partial \log{\mu}}  = - \frac{29}{5\pi} (\Lambda G)^2,
\ee
which determines the RG flow of the dimensionless quantity $\Lambda G$. Interestingly, the $\beta$-function (\ref{RG-gr}) is {\it negative}, which means that the currently measured exceedingly small value of $\Lambda G\sim 10^{-120}$ decreases even further for higher energies. Of course, this running is extremely slow, first because it is only logarithmic, and second because it is controlled by $(\Lambda G)^2$.

The interpretation is rather different in the pure connection formulation.
Here, for a general member of the class of theories (\ref{action}), the action evaluated on an instanton (\ref{X-inst}) becomes
\be
S[{\rm instanton}] = - 2\left(\frac{\Lambda}{3}\right)^2 f(\delta) \int d^4x\sqrt{g}.
\ee
Equating its scale derivative with the volume term in (\ref{result}), we find the $\beta$-function for $f(\d)$
\be\label{RG-conn}
\frac{\partial f(\delta)}{\partial \log(\mu)} = \frac{121}{5(4\pi)^2},
\ee
after cancelling the factors of $\Lambda^2$ and the volume on both sides.

However, given that the background used corresponds to $\tilde{X}^{ij}\sim\delta^{ij}$, the flow (\ref{RG-conn}) only contains information about the running of the value of the function $f$ at a single point.
Therefore, no information about how the shape of the function $f$ changes with energy is contained in (\ref{RG-conn}).
For instance, our calculation does not allow to distinguish a term in $f$ proportional to ${\rm Tr}(\tilde{X})$ from a term proportional to $({\rm Tr}\sqrt{\tilde{X}})^2$, since on $\tilde{X}^{ij}\sim\delta^{ij}$ both reduce to just a number. While the former is a topological term that can always be added without changing the field equations, the coefficient of the latter encodes the $\beta$-function for $1/\Lambda G$ in view of (\ref{f-GR}).
Thus, at present, we cannot even determine the sign of the $\beta$-function for $\Lambda G$ in the connection formulation. Our result is thus just a first step towards more general understanding.
In particular, it can be used as a consistency check, as the flow for an arbitrary function $f$ on a more general background has to reduce to (\ref{RG-conn}) when evaluated on the identity matrix.
We hope to return to the more general problem in the future.

In this paper we considered only the case of pure gravity, not including any coupling to matter.
In contrast to the metric formulation, in the pure connection formulation, one cannot couple matter in the usual way, as no metric is available.
Instead, in the context of diffeomorphism invariant gauge theories one can couple matter only by enlarging the gauge group of the theory, see \cite{Krasnov:2011hi} for a description of this idea.
Many types of matter can be added this way, e.g. gauge fields with Yang-Mills dynamics, as well as scalar and higher spin fields.
It would be interesting to generalize the calculations in this paper for a theory of the type (\ref{action}) with a larger gauge group.
We hope to approach this investigation in the future.

\section*{Acknowledgements} The authors were supported by an ERC Starting Grant 277570-DIGT. The second author also acknowledges support from the Alexander von Humboldt foundation, Germany, and from Max Planck Institute for Gravitational Physics, Golm. KK is grateful to Joel Fine for correspondence on some aspects of instantons, as well as for some suggestions that led to improvements in the second version. KK would also like to thank Dario Benedetti for pointing out a reference and Dima Vassilevich for an email exchange about the zero modes.

\section*{Appendix: One loop relation between the metric and connection formulations}

The purpose of this Appendix is to provide an explanation for the observed close relationship between the one loop results in metric and connection formulations. Indeed, we have started this article by stating that theories that are only equivalent on shell can in principle lead to different quantum theories. Here we explain why one finds little difference at the one loop level.

We start with the Plebanski action functional, from which both the metric and connection descriptions can be obtained by integrating out fields. This functional reads
\be\label{action-Pleb}
S[A,\Sigma,\Psi] = \frac{1}{8\pi G} \int \Sigma^i\wedge F^i - \frac{1}{2} \left( \Psi^{ij} - \frac{\Lambda}{3} \delta^{ij}\right) \Sigma^i\wedge \Sigma^j.
\ee
Here $\Sigma^i$ are ${\mathfrak so}(3)$-valued 2-forms, and $\Psi^{ij}$ is a symmetric $3\times 3$ matrix with zero trace. As before $A^i$ is an ${\rm SO}(3)$ connection and $F^i$ is its curvature. When $\Psi^{ij}$ is varied one obtains $\Sigma^i\wedge \Sigma^j\sim\delta^{ij}$. Introducing the metric that makes $\Sigma^i$ into self-dual forms, and has the volume form $({\rm vol})=(1/6) \, \Sigma^i\wedge \Sigma^i$, one can interpret the other Euler-Lagrange equations of this theory as Einstein equations for this metric. Thus, the equation $d_A \Sigma^i=0$ obtained by varying the Lagrangian with respect to the connection can be shown to imply that $A^i$ is the self-dual part of the Levi-Civita connection compatible with the metric. Then $F^i=(\Psi^{ij} - (\Lambda/3)\delta^{ij})\Sigma^j$ is the statement that the curvature of the self-dual part of the Levi-Civita connection is self-dual, which is the Einstein condition. One can also show that when the action (\ref{action-Pleb}) is evaluated on the configuration $\Sigma^i\wedge \Sigma^j\sim\delta^{ij}$ with $d_A \Sigma^i=0$, it reduces to the Einstein-Hilbert action (for the metric defined by $\Sigma^i$), plus a topological term. This is another way of saying that when the fields $A^i, \Psi^{ij}$ are integrated out, one reobtains the Einstein-Hilbert action. For more details about the Plebanski formulation, the reader can consult e.g. \cite{Krasnov:2009pu}.

We proceed by linearising the Plebanski action around an arbitrary background. From the description in the previous paragraph it is clear that if we then integrate out the linearised fields $a^i, \psi^{ij}$, we will obtain the linearised Einstein-Hilbert action. This shows that the one loop calculation based on the Plebanski action will produce the same result as in the standard metric formulation. Then, we can instead integrate out the fields $b^i,\psi^{ij}$, and obtain the linearised Lagrangian of the connection formulation. Here we show how this leads to the linearised Lagrangian (\ref{lin-L*}) when an instanton background is chosen. This argument explains why the metric and connection calculations lead to so closely related results.

The calculation proceeds as follows. Replacing everywhere $\Sigma^i\to \Sigma^i + b^i, A^i\to A^i+a^i, \Psi^{ij}\to \Psi^{ij} + \psi^{ij}$ we get the following terms quadratic in the perturbations:
\be
8\pi G {\cal L}^{(2)}= \frac{1}{2} \Sigma^i \wedge [a,a]^i + b^i \wedge d_A a^i - \psi^{ij} \Sigma^i \wedge b^j -\frac{1}{2} \left( \Psi^{ij} - \frac{\Lambda}{3}\delta^{ij}\right) b^i \wedge b^j .
\ee
Our aim is to show how (\ref{lin-L*}) results from this. Let us first integrate out $b^i$ via its field equation
\be\label{app-b}
b^i = \left(\Psi^{ij}- \frac{\Lambda}{3}\delta^{ij}\right)^{-1} \left( d_A a^j - \psi^{jk} \Sigma^k \right),
\ee
where a matrix inverse appears. We find
\be\label{app-1}
8\pi G {\cal L}^{(2)} = \frac{1}{2} \Sigma^i \wedge [a,a]^i  + \frac{1}{2} \left(\Psi^{ij}- \frac{\Lambda}{3}\delta^{ij}\right)^{-1}  \left(d_A a^i- \psi^{im} \Sigma^m \right)\wedge \left(d_A a^j - \psi^{jn} \Sigma^n \right).
 \ee
The second step is to integrate out $\psi^{ij}$. To solve the arising equations, let us represent the self-dual part of the two-form $d_A a^i$ as
 \be\label{app-M}
 (d_A a^i)_{sd} = M^{ij} \Sigma^j.
 \ee
Thus we have
 \be\label{app-2}
 \left(\Psi^{ik}- \frac{\Lambda}{3}\delta^{ik}\right)^{-1} (M^{kj}- \psi^{kj}) \Big|_{tf} = 0,
 \ee
 where the projection on the $ij$ symmetric tracefree part is taken.

It is at this step that specialising to an instanton background leads to great simplifications. On this background $\Psi^{ij}=0$ and the solution of (\ref{app-2}) is simply
 \be
 \psi^{ij} = M^{(ij)}_{tf}=\frac{1}{4} P^{ij|kl} (\Sigma^{k\mu\nu} d_\mu a_\nu^l),
 \ee
 i.e.~the symmetric tracefree part of $M^{ij}$. The second expression gives this matrix explicitly. This result means that in each of the brackets in the second term in (\ref{app-1}), the symmetric tracefree part of the self-dual part of $d_A a^i$ gets cancelled by $\psi^{ij}$, while all other parts of $d_A a^i$ remain. To see what the action reduces to in this case, we use $\Sigma^i=-(3/\Lambda) F^i$ to rewrite the first term as
 \be\label{app-3}
- \frac{3}{2\Lambda} F^i \wedge \epsilon^{ijk} a^j\wedge a^k = \frac{3}{2\Lambda} a^i\wedge d_A d_A a^i = \frac{3}{2\Lambda} d_A a^i\wedge d_A a^i ,
\ee
where we have integrated by parts to get the last equality. The linearised Lagrangian therefore becomes
\be
 {\cal L}^{(2)} = \frac{3M_p^2}{\Lambda} \left( d_A a^i\wedge d_A a^i - \left( d_A a^i - \widetilde{d_A a^i}\right) \wedge \left( d_A a^i - \widetilde{d_A a^i}\right)\right),
 \ee
 where we have introduced $M_p^2=1/16\pi G$ and $\widetilde{d_A a^i}$ is the projection of the self-dual part of $d_A a^i$ which is captured by the symmetric tracefree part of the matrix $M^{ij}$, i.e.
 \be\label{app-4}
 \widetilde{d_A a^i} := \frac{1}{4} P^{ij|kl} (\Sigma^{k\mu\nu} d_\mu a_\nu^l) \Sigma^j.
\ee
All terms in the brackets now cancel each other, except for the $\widetilde{d_A a^i}$ terms. So, we finally arrive at
 \be\label{app-final}
  {\cal L}^{(2)} = \frac{3M_p^2}{\Lambda} \left(  \widetilde{d_A a^i} \wedge \widetilde{d_A a^j}\right),
 \ee
 which, using (\ref{app-4}) can be seen to coincide with (\ref{lin-L*}). Here we have used the fact that $\widetilde{d_A a^i}$ is self-dual, and so is orthogonal to the anti-self-dual parts of $d_A a^i$. It is also orthogonal to the other self-dual parts, and so we have $\widetilde{d_A a^i}\wedge \left( d_A a^i - \widetilde{d_A a^i}\right) =0$, which was used to get (\ref{app-final}).

This calculation explains why the one loop results in the metric and connection formulations are so closely related. We have spelled out the details of this argument only for an instanton background, in which case the calculation simplifies. It is also clear that the same argument establishes that one loop results in the metric and connection formulations for an arbitrary background will coincide (modulo subtleties related to the scalar mode). However, from the calculation of the previous paragraph it is clear that for an arbitrary background the pure connection linearised Lagrangian is much more involved than in the instanton case. We will not attempt the general background connection formulation calculation in this paper. However, it is important to know that we should anticipate a result closely related to the one obtained in the metric GR.

The arguments above also tell us that {\it beyond} one loop no simple relation between the metric and connection based quantum theories should be expected. This has to do with non-linearities of the field equations that are solved with solutions substituted into the action in the passage from Plebanski and pure connection formulations. This passage is explained in more details in \cite{Krasnov:2011up}. In the first step one solves the linear field equation for the $\Sigma^i$ fields. However, in the next step one has to solve a very non-linear equation for the $\Psi^{ij}$ matrix. It is at this step that the square root of the matrix $F^i\wedge F^j$ appears. Once this equation is solved and the solution for $\Psi^{ij}$ is substituted into the action, some information about the off-shell behaviour of the Plebanski action (\ref{action-Pleb}) is lost. This is why the pure connection description is only on-shell equivalent to (\ref{action-Pleb}) and thus to the metric GR. At the one loop level this does not lead to any differences in the results (once again apart from the subtleties of the scalar modes). But at higher loops such differences may well arise. This is interesting as we know that the vacuum GR only exhibits non-renormalisability at the two loop order \cite{Goroff:1985th}. It would be very interesting to see what are the two loop properties of the pure connection version of GR, but this is beyond the scope of this paper.

\eject

\end{document}